\documentclass[reprint,superscriptaddress,amsmath,amssymb,aps,prd,longbibliography,floatfix,nofootinbib]{revtex4-1}

\usepackage{tensor}     
\usepackage{graphicx}   
\usepackage[
colorlinks=true,        
citecolor=blue,         
linkcolor=blue,         
urlcolor=blue           
]{hyperref}             
\usepackage{bm}         
\usepackage{xcolor}     
\usepackage{color}      
\usepackage[utf8]{inputenc} 
\usepackage[section]{placeins} 
\DeclareUnicodeCharacter{2009}{\,}
\newcommand{\nc}{\newcommand*} 

\nc{\al}{\alpha}
\nc{\s}{\sigma}
\nc{\dt}{\delta}
\nc{\Dt}{\Delta}
\nc{\Ld}{\Lambda}
\nc{\p}{\partial}
\nc{\Om}{\Omega}
\nc{\rd}{\mathrm{d}}
\nc{\Od}{\mathcal{O}} 

\def\({\left(}
\def\){\right)}
\def\[{\left[}
\def\]{\right]}
\def\e{\begin{equation}}
\def\q{\end{equation}}
\def\m{\begin{eqnarray}}
\def\n{\end{eqnarray}}

\nc{\Eq}[1]{Eq.~\eqref{#1}}     
\nc{\Fig}[1]{Fig.~\ref{#1}}     
\nc{\Table}[1]{Table~\ref{#1}}  
\nc{\Sec}[1]{Sec.~\ref{#1}}     

\nc{\Msun}{M_\odot}             
\nc{\fpbh}{f_{\mathrm{PBH}}}    
\nc{\fpbhn}{f_{\mathrm{PBH0}}}    
\nc{\mR}{\mathcal{R}} 
\nc{\seq}{\sigma_{\mathrm{eq}}}
\nc{\ogw}{\Omega_{\mathrm{GW}}}
\nc{\gpcyr}{\mathrm{Gpc}^{-3}\,\mathrm{yr}^{-1}}
\nc{\lvc}{LIGO/Virgo} 
\nc{\SNR}{\mathrm{SNR}} 
\nc{\mmin}{{m_{\mathrm{min}}}}
\nc{\mmax}{{m_{\mathrm{max}}}}
\nc{\Mmin}{{M_{\mathrm{min}}}}
\nc{\fmin}{{f_{\mathrm{min}}}}
\nc{\VT}{\mathrm{VT}}
\nc{\rhoGW}{\rho_{\mathrm{GW}}}
\nc{\vth}{\vec{\theta}}
\nc{\vd}{\vec{d}}
\nc{\vla}{\vec{\lambda}}
\nc{\Nobs}{N_{\mathrm{obs}}}
\nc{\av}[1]{\langle #1 \rangle} 
\nc{\km}{\mathrm{km}}
\nc{\Mpc}{\mathrm{Mpc}}
\nc{\Tobs}{T_{\mathrm{obs}}}
\nc{\Ntemp}{N_{\mathrm{temp}}}

\nc{\addref}{[\textcolor{red}{add ref}] } 
\nc{\eg}{\textit{e.g.~}}
\nc{\app}{\approx}
\nc{\hf}{\frac{1}{2}}
\nc{\discuss}{\textcolor{red}{Add discussion here!}}
\nc{\red}[1]{\textcolor{red}{#1}}

\begin{document}

\title{Distinguishing Primordial Black Holes from Astrophysical Black Holes by Einstein Telescope and Cosmic Explorer}

\author{Zu-Cheng Chen}
\email{chenzucheng@itp.ac.cn} 
\affiliation{CAS Key Laboratory of Theoretical Physics, 
Institute of Theoretical Physics, Chinese Academy of Sciences,
Beijing 100190, China}
\affiliation{School of Physical Sciences, 
University of Chinese Academy of Sciences, 
No. 19A Yuquan Road, Beijing 100049, China}

\author{Qing-Guo Huang}
\email{huangqg@itp.ac.cn}
\affiliation{CAS Key Laboratory of Theoretical Physics, 
Institute of Theoretical Physics, Chinese Academy of Sciences,
Beijing 100190, China}
\affiliation{School of Physical Sciences,
University of Chinese Academy of Sciences, 
No. 19A Yuquan Road, Beijing 100049, China}
\affiliation{Center for Gravitation and cosmology, 
College of Physical Science and Technology, Yangzhou University, 
88 South University Ave., 225009, Yangzhou, China}
\affiliation{Synergetic Innovation Center for Quantum Effects and Applications, 
Hunan Normal University, 36 Lushan Lu, 410081, Changsha, China}

\begin{abstract}

We investigate how the next generation gravitational-wave (GW) detectors, such as Einstein Telescope (ET) and Cosmic Explorer (CE), can be used to distinguish primordial black holes (PBHs) from astrophysical black holes (ABHs). 
Since a direct detection of sub-solar mass black holes can be taken as the smoking gun for PBHs, we estimate the detectable limits of the abundance of sub-solar mass PBHs in cold dark matter by the targeted search for sub-solar mass PBH binaries and binaries containing a sub-solar mass PBH and a super-solar mass PBH, respectively. 
On the other hand, according to the different redshift evolutions of the merger rate for PBH binaries and ABH binaries, we forecast the detectable event rate distributions for the PBH binaries and ABH binaries by ET and CE respectively, which can serve as a method to distinguish super-solar mass PBHs from ABHs.

\end{abstract}

\maketitle

\section{\label{intro}Introduction}
Ten binary black hole (BBH) mergers were detected during \lvc\ O1 and O2 observing runs \cite{Abbott:2016blz,Abbott:2016nmj,TheLIGOScientific:2016pea,Abbott:2017vtc,%
Abbott:2017gyy,Abbott:2017oio,LIGOScientific:2018mvr}.
Understanding the origin of these BBHs is an essential scientific goal, which is still under intensive investigation (see \eg \cite{Bird:2016dcv,Sasaki:2016jop,Chen:2018czv,Clesse:2017bsw,Fishbach:2017dwv,Clesse:2016vqa,Antonini:2016gqe,Inayoshi:2017mrs,Ali-Haimoud:2017rtz,Perna:2019axr,Kavanagh:2018ggo,Rodriguez:2015oxa,Rodriguez:2016kxx,Park:2017zgj,Wu:2020drm,Belczynski:2014iua,Belczynski:2016obo,Woosley:2016nnw,Rodriguez:2018rmd,Choksi:2018jnq,2010AIPC.1314..291D,deMink:2016vkw}).
The fact that the component masses of these BBHs observed by gravitational waves (GWs) exhibit a much heavier mass distribution than the one inferred from X-ray observations \cite{Wiktorowicz:2013dua,Casares:2013tpa,Corral-Santana:2013uua,%
Corral-Santana:2015fud} has triggered the interest in the community to speculate that the observed BBH mergers might come from the stellar mass primordial black holes (PBHs) \cite{Bird:2016dcv,Sasaki:2016jop,Chen:2018czv,Clesse:2017bsw}.

PBHs are the black holes (BHs) that form in the early universe by gravitational collapse of primordial density perturbations \cite{Hawking:1971ei,Carr:1974nx,Khlopov:2008qy,Sasaki:2018dmp} and undergo quite different evolutionary histories than the astrophysical black holes (ABHs) which originate from the demise of massive stars.
PBHs may contribute to a fraction of cold dark matter (CDM), and the abundance of PBHs in CDM, $\fpbh$, has been constrained by a variety of experiments, \eg extra-galactic gamma-ray \cite{Carr:2009jm}, femtolensing of gamma-ray bursts \cite{Barnacka:2012bm},
the existence of white dwarfs in our local galaxy \cite{Graham:2015apa}, Subaru/HSC microlensing \cite{Niikura:2017zjd}, Kepler milli/microlensing \cite{Griest:2013esa}, OGLE microlensing \cite{Niikura:2019kqi}, EROS/MACHO microlensing \cite{Tisserand:2006zx,Calcino:2018mwh}, dynamical heating of ultra-faint dwarf galaxies \cite{Brandt:2016aco}, X-ray/radio constraints \cite{Gaggero:2016dpq}, cosmic microwave background (CMB) spectrum \cite{Ali-Haimoud:2016mbv,Blum:2016cjs,Horowitz:2016lib,Chen:2016pud,Poulin:2017bwe}, and GWs \cite{Wang:2016ana,Abbott:2018oah,Authors:2019qbw,Magee:2018opb,Wang:2019kaf,Chen:2019xse,Yuan:2019udt,Chen:2018rzo}.

An alternative way to explain \lvc\ BBHs is through ABH models.
The formation and merger of ABHs are guided by evolutionary environments.
There exist three main channels in the literature.
The first one is the \textit{dynamical formation} channel, in which BHs are formed through the evolution of massive stars and segregated to the cluster core to pair as BBHs \cite{Rodriguez:2015oxa,Rodriguez:2016kxx,Park:2017zgj}. 
The second one is the \textit{classical isolated binary evolution} channel, in which the BBHs are formed through highly non-conservative mass transfer or common envelope ejection \cite{Belczynski:2014iua,Belczynski:2016obo,Woosley:2016nnw,Rodriguez:2018rmd,Choksi:2018jnq}.
The third one is the \textit{chemically homogeneous evolution}, in which stars evolve almost chemically homogeneously to form BHs because of the mixing of helium produced in the center throughout the envelope \cite{2010AIPC.1314..291D,deMink:2016vkw}.
Properties of BBHs, such as the spin \cite{Farr:2017uvj,Tiwari:2018qch,Ng:2018neg,Stevenson:2017dlk,Bogomazov:2018prw,Lopez:2018nkj,Sedda:2018nxm,Farr:2017gtv}, redshift \cite{Fishbach:2018edt,Emami:2018taj,Bai:2018shq}, and eccentricity distributions \cite{Samsing:2013kua,Samsing:2017xmd,Samsing:2017jnz,Lower:2018seu} have been proposed to discriminate different channels of astrophysical origin BBH (AOBBH) models.


In this paper, we will explore and forecast the possibility of distinguishing PBHs from  ABHs by using GW observations, especially by the third generation ground-based GW detectors like Einstein Telescope (ET) \cite{Punturo:2010zz} and Cosmic Explorer (CE) \cite{Evans:2016mbw}, which are expected to detect many more BBHs than current \lvc, at an order of $\Od(10^5)$ events per year \cite{Regimbau:2016ike,Vitale:2018yhm}. 
The rest of this paper is organized as follows.
In \Sec{subsolar}, we focus on the sub-solar mass ($\lesssim 1\Msun$) BBHs.
Because ABHs are expected to be heavier than the Chandrasekhar mass limit $\sim 1.4 \Msun$ \cite{Chandrasekhar:1931ftj,Chandrasekhar:1931ih}, direct detection of sub-solar mass BHs can be evidence of PBHs.
In \Sec{mono}, assuming PBHs have a monochromatic mass distribution, we estimate the detectable limits on the abundance of PBHs from the targeted search by ET and CE, respectively. 
In \Sec{general}, considering that PBHs have a broad mass distribution and all the BBH events detected by LIGO/Virgo originate from PBHs, we adopt a model-independent approach to constrain the abundance of PBHs with super-solar mass ($\gtrsim 1\Msun$) from \lvc\ events, and then explore the detectable limits on the abundance of sub-solar mass PBHs by searching for the BBHs containing a sub-solar mass PBH and a super-solar mass PBH. 
\Sec{stellar} is dedicated to the super-solar mass BBHs.
The redshift evolution of the merger rate for primordial origin BBHs (POBBHs) and AOBBHs can be quite different, which results in different redshift distributions of the expected number of observable BBHs.
We estimate and forecast the event number distributions of the PBH and ABH models for ET and CE respectively, which can serve as a complementary tool to distinguish PBHs from ABHs.
Finally, we summarize and discuss our results in \Sec{discuss}

\section{\label{subsolar}Distinguish PBHs from ABHs by Sub-solar Mass BHs}	
Direct detection of sub-solar mass BHs can be taken as a smoking gun for PBHs.
Nonetheless, the event rate relies both on the merger rate of POBBHs and the sensitivity of GW detectors.
In the following two subsections, we will explore the abilities to detect sub-solar mass BHs for different GW detectors by considering the cases when PBHs have a monochromatic and a general mass functions, respectively.

\subsection{\label{mono}Monochromatic mass function}
In this subsection, we assume all the PBHs have the same mass and estimate the detectable limits of $\fpbh$ by the targeted search of POBBHs.


The redshift $z$ evolution of the local merger rate $R(z)$ in the comoving frame for the monochromatic mass function, by accounting the angular momentum exerted both by all PBHs and the background inhomogeneity, is given by \cite{Ali-Haimoud:2017rtz,Chen:2018czv}
\e\label{mono_R} 
	R(z) = 3.9 \times 10^6 \times \({\frac{t(z)}{t_0}}\)^{-\frac{34}{37}}
		 m^{-32/37} f^2 \(f^2 + \seq^2\)^{-21/74},
\q 
where $m \Msun$ is the component mass of BBHs measured in the source frame, and $\sigma_{\mathrm{eq}}$ is the variance of density perturbations of the rest DM on scale of order $\Od(10^0\sim10^3) M_\odot$ at radiation-matter equality. 
Following \cite{Ali-Haimoud:2017rtz,Chen:2018czv}, we choose $\sigma_{\mathrm{eq}}\approx 0.005$.
Here $\fpbh \equiv \Omega_{\mathrm{PBH}}/\Omega_{\mathrm{CDM}}$ is the energy density fraction of PBHs in CDM and is related to the total abundance of PBHs in non-relativistic matter, $f$, by $\fpbh \approx f/0.85$. 
Besides, $t(z)$ is the cosmic time at the redshift of $z$ and $t_0 \equiv t(0)$ is the age of our universe.
Note that throughout this paper, we adopt the units in which the Newtonian constant $G$ and the speed of light in vacuum $c$ are set to unity.
Here we ignore the effects of early formed PBH clusters which may alter the merger rate of PBH binaries and present a rich phenomenology \cite{Vaskonen:2019jpv,Trashorras:2020mwn,Jedamzik:2020ypm}. We hope to come back to this issue in the future.

\begin{figure}[htbp!]
    \centering
    \includegraphics[width=0.48\textwidth]{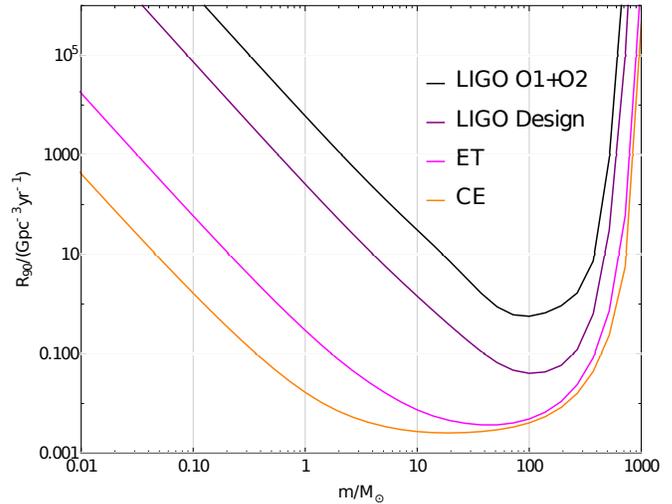}
    \caption{\label{R90_plot}The $90\%$ confidential upper limit on the binary merger rate, $R_{90}$, as a function of the masses of the equal-mass BBHs for LIGO O1 \& O2, LIGO Design, ET and CE.}
\end{figure}

The expected number of detection, $\Nobs$, then follows \cite{Chen:2018czv,Kavanagh:2018ggo}
\e\label{Nobs} 
	\Nobs = \int R(z) \frac{\rd VT}{\rd z} \rd z,
\q 
where $\rd VT/\rd z$ is the spacetime sensitivity of a GW detector as a function of redshift and accounts for the selection effects of that detector.
Generally, $\rd VT/\rd z$ depends on the properties $\xi$ (\eg  masses and spin) of a binary and is defined as \cite{Abbott:2016nhf,Abbott:2016drs}
\e 
	\frac{\rd VT}{\rd z} = \frac{\rd V_c}{\rd z} \frac{\Tobs}{1+z} f(z|\xi),   
\q 
where $V_c$ is the comoving volume \cite{Hogg:1999ad}, $\Tobs$ is the observing time, and the denominator $1+z$ accounts for the converting of cosmic time from source frame to detector frame due to the cosmic redshift.
Here $0 < f(z|\xi) < 1$ is the probability of detecting a BBH with the given parameters $\xi$ at redshift $z$ \cite{OShaughnessy:2009szr}.
The $90\%$ confidential upper limit on the binary merger rate can then be obtained by using the loudest event statistic formalism \cite{Biswas:2007ni},
\e\label{R90} 
R_{90} = \frac{2.303}{VT},
\q
where 
\e\label{VT}
VT = \int \frac{\rd VT}{\rd z} \rd z.
\q
We adopt the semi-analytical approximation from \cite{Abbott:2016nhf,Abbott:2016drs} to calculate $VT$ by neglecting the effect of spins for BHs and using the \texttt{IMRPhenomPv2} waveform to simulate the BBH templates.
Furthermore, we set a single-detector signal-to-noise ratio (SNR) threshold $\rho_{\mathrm{th}} = 8$ as a criterion of detection, which roughly corresponds to a network threshold of $12$.
\Fig{R90_plot} shows the estimated $90\%$ upper limits on the merger rate of equal-mass BBHs, $R_{90}$, which are estimated from \Eq{R90}.

\begin{figure*}[t]
	\centering
	\includegraphics[width=\textwidth]{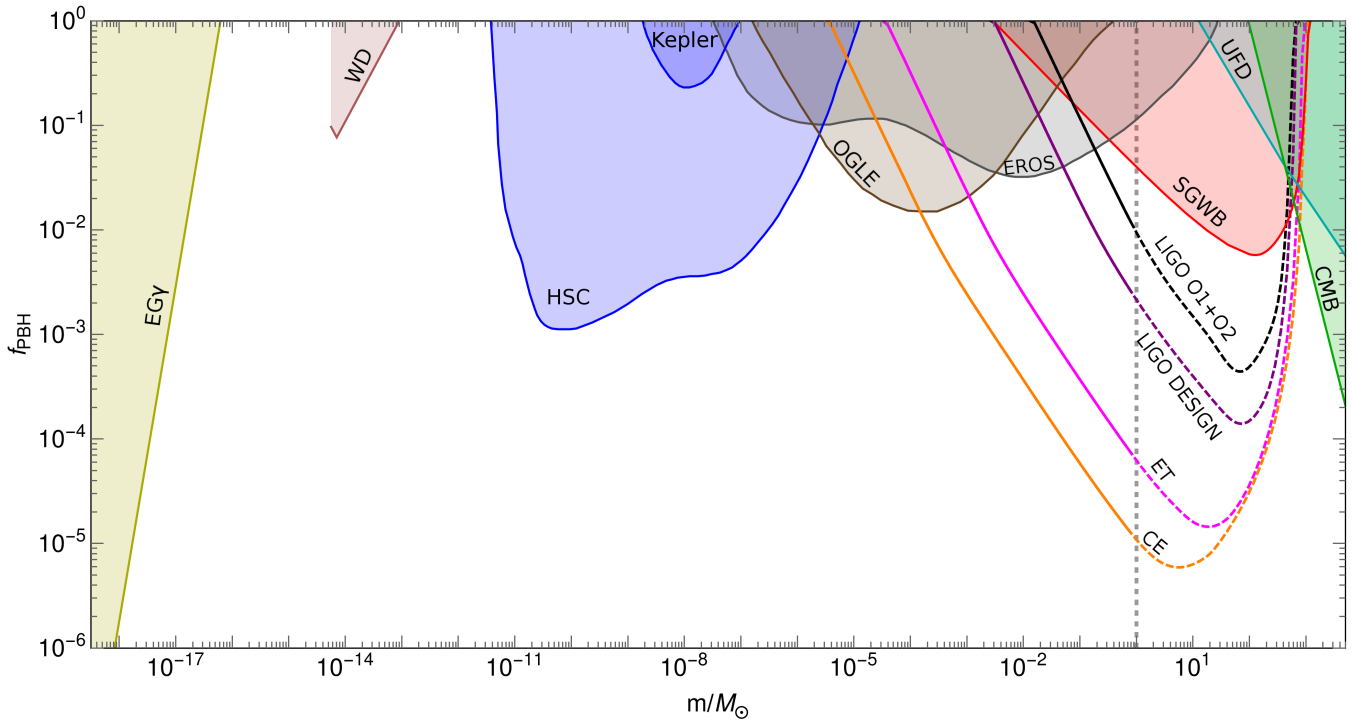}
	\caption{\label{fpbh} Constraints on the abundance of PBHs, $\fpbh$, 
		with a monochromatic mass distribution both by the non-detection of SGWBs
		and the null targeted search result of BBHs.
		The gray vertical line at $1\Msun$ indicates that the constraints from
		the targeted search are only valid for the sub-solar mass PBHs, 
		because we yet cannot conclude that none of the ten BBHs detected by 
		\lvc\ are of POBBHs. 
		The black, purple, magenta and orange curves are the results of targeted
		search from LIGO O1 \& O2, LIGO Design, ET and CE, respectively.
		The observing times of LIGO Design, ET and CE are all assumed 
		to be $1$ year.
		The red curve is the updated upper bound of $\fpbh$ constrained by 
		the non-detection of SGWB from both LIGO O1 and O2 searches.
		The results from other experiments are also shown here:
		extra-galactic gamma-ray (EG$\gamma$) \cite{Carr:2009jm},
		existence of white dwarfs in our local galaxy (WD) \cite{Graham:2015apa}, 
		Subaru HSC microlensing (HSC) \cite{Niikura:2017zjd}, 
		Kepler milli/microlensing (Kepler) \cite{Griest:2013esa}, 
		EROS/MACHO microlensing (EROS) \cite{Tisserand:2006zx}, 
		OGLE microlensing (OGLE) \cite{Niikura:2019kqi},
		dynamical heating of ultra-faint dwarf galaxies (UFD)%
		\cite{Brandt:2016aco},
		and accretion constraints by CMB \cite{Ali-Haimoud:2016mbv,Blum:2016cjs,%
		Horowitz:2016lib,Chen:2016pud,Poulin:2017bwe}.
	}
\end{figure*}

The detectable limit of $\fpbh$ by the targeted search from LIGO O1 \& O2, LIGO Design, ET and CE are shown in \Fig{fpbh}.
The effective observing time of LIGO O1 \& O2 is set to their total running time of $165.6$ days \cite{TheLIGOScientific:2016pea,TheLIGOScientific:2017qsa}.
Meanwhile LIGO Design, ET and CE are supposed to operate at $1$ year with full duty.
The upper limit for $\fpbh$ of sub-solar mass PBHs in the mass range $\[0.2, 1\]\Msun$ has been reported in \cite{Abbott:2018oah,Magee:2018opb} by the null targeted search result of BBHs in that mass range.
We extrapolate the results of \cite{Abbott:2018oah,Magee:2018opb} in several aspects.
Firstly, we adopt the merger rate presented in \cite{Ali-Haimoud:2017rtz}, which takes a more careful examination of the dynamical evolution of the binary systems than the one used by \cite{Abbott:2018oah,Magee:2018opb} from the results of \cite{Sasaki:2016jop}.
Secondly, we also estimate the detectable limits of $\fpbh$ by the proposed third generation GW detectors like CE and ET.
Lastly, we do not limit to the mass range of $\[0.2, 1\]\Msun$ but extend to the range constrained by the detectors automatically.
In particular, since whether the super-solar mass BBHs observed so far are of POBBHs or not is still under debate, we use the dashed lines to illustrate the detectable limits of $\fpbh$ for the super-solar mass PBHs. 
Furthermore, as the masses go down below $0.2 \Msun$, the search difficulty arises because the number of templates required in the template bank, $\Ntemp$, scales both as the minimum mass $\Mmin$ and the starting frequency $\fmin$ \cite{Magee:2018opb}
\e 
	\Ntemp \propto \(\Mmin  \fmin\)^{-8/3}.
\q
The dramatic increment of computational resource limits the current GW search pipeline that it cannot deal with the BBHs with component masses far below $0.2 \Msun$ efficiently.
However, in the future, this issue may be addressed by the improvement of search algorithms or computational technologies.
Note also that the upper limits of $\fpbh$ in mass range of $[10, 300]\Msun$ has already been derived in \cite{Ali-Haimoud:2017rtz,Kavanagh:2018ggo} by using LIGO/Virgo O1 data.
We updated their constraints by using both O1 and O2 data. And therefore our upper limits are more stringent than those in \cite{Ali-Haimoud:2017rtz,Kavanagh:2018ggo}.

Besides, we update the constraint on $\fpbh$ from the null search of stochastic gravitational-wave background (SGWB) from LIGO O1 in \cite{Wang:2016ana} to both LIGO O1 and O2 runs \cite{TheLIGOScientific:2016dpb,Renzini:2018nee,LIGOScientific:2019vic,LIGOScientific:2019gaw}. 
The method adopted in this paper is described in the Appendix. 
Usually one may expect that the null detection in the targeted search would give a tighter constraint on the abundance of PBHs than that from SGWBs. 
However, the GW signals from light BBHs are so weak to be resolved individually by GW observations, and hence the null detection of SGWBs provides a more stringent constraint on the abundance of light PBHs because those weak signals can superpose to form a detectable SGWB. 
See a cross of the red and black curves around $0.1 M_\odot$ in \Fig{fpbh}. 
Actually, it is also true for other detections, such as LIGO design, ET and CE. 

\subsection{\label{general}General mass function -- a model independent approach}

A null search result of the sub-solar mass PBHs in LIGO's O1 and O2 data has been reported in \cite{Abbott:2018oah,Magee:2018opb,Authors:2019qbw}.
However, the search was only targeted at the POBBHs with component masses between $0.2\Msun$ and $1\Msun$.
In this subsection, we propose to search for the POBBHs with a component of sub-solar mass and the other of super-solar mass, which are expected to emit stronger GWs. 

Here we extend the discussion in the former subsection to the case in which PBHs exhibit a broad mass distribution, and assume that all the ten BBHs observed by \lvc\ so far are POBBHs. 
Contrary to the previous works \cite{Chen:2018rzo,Raidal:2017mfl,Kocsis:2017yty,Raidal:2018bbj} by choosing some specific mass functions, \eg a power-law or lognormal ones, which are pertinent to some specific formation models of PBHs, we take a model-independent approach by binning the mass function $P(m)$ from $0.2 M_\odot$ to $100 M_\odot$ as
\e\label{para} 
P(m) = \begin{cases} 
	P_0, & 0.2\, \Msun \leq m < 1\, \Msun \\
	P_1, & 1\, \Msun \leq m < 30\, \Msun \\
	P_2, & 30\, \Msun \leq m < 60\, \Msun \\
	P_3, & 60\, \Msun \leq m \leq 100\, \Msun
\end{cases}
\q 
where $P_i = \{ P_0, P_1, P_2, P_3 \} $ are four constants satisfying the normalization condition
\e 
\int P(m)\, \rd m = 0.8P_0 + 29P_1 + 30P_2 + 40P_3 = 1.
\q 
Here, only three out of the four $P_i$s are independent and we will choose $\vth = \{P_1, P_2, P_3\}$ as free parameters.
$\vth$ will be fitted by the ten BBHs from LIGO's O1 and O2 runs. 
In this subsection, we are only interested in the PBHs with masses in the range of $\[\mmin, \mmax\] = \[0.2\Msun, 100\Msun\]$.
Here $\mmin = 0.2\Msun$ corresponds to the lower mass bound of LIGO's O1 search for the sub-solar mass ultracompact binaries in \cite{Abbott:2018oah}.

The merger rate for POBBHs with a general mass function is given in \cite{Chen:2018czv}.\footnote{In \cite{Raidal:2017mfl,Kocsis:2017yty,
Liu:2019rnx,Liu:2018ess,Raidal:2018bbj}, the authors ignored the relative distribution of PBHs, and cannot guarantee that PBH binaries be formed from the closest neighboring PBHs.}
The time-dependent comoving merger rate density for a general normalized mass function, $P(m|\vth)$, takes the form of
\m\label{calR} 
\mR_{12}&&(t|\vth) \app 3.9 \cdot 10^6 \times \({\frac{t}{t_0}}\)^{-\frac{34}{37}} f^2 (f^2+\sigma_{\mathrm{eq}}^2)^{-{21\over 74}} \nonumber \\
&& \times  \min\(\frac{P(m_1|\vth)}{m_1}, \frac{P(m_2|\vth)}{m_2}\) \({P(m_1|\vth)\over m_1}+{P(m_2|\vth)\over m_2}\) \nonumber \\
&& \times (m_1 m_2)^{{3\over 37}} (m_1+m_2)^{36\over 37},
\n
in units of $\gpcyr$, where the component masses $m_1$ and $m_2$ are in units of $\Msun$.
The time (or redshift)-dependent merger rate can be obtained by integrating over the component masses
\e\label{Rcal2}
	\mR(t|\vth) = \int \mR_{12}(t|\vth)\ \rd m_1\, \rd m_2.
\q 
The local merger rate density distribution then reads \cite{Chen:2018rzo}
\e 
\mR_{12}(t_0|\vth) = R\, p(m_1,m_2|\vth),
\q 
where the local merger rate $R \equiv \mR(t_0|\vth)$ is chosen such that the population distribution of BBH mergers, $p(m_1,m_2|\vth)$, is normalized. 

To extract the population parameters $\{\vth, R\}$ from the merger events observed by \lvc, it is necessary to perform the hierarchical Bayesian inference on the BBHs' mass distribution \cite{Abbott:2016nhf,Abbott:2016drs,TheLIGOScientific:2016pea,Wysocki:2018mpo,Fishbach:2018edt,Mandel:2018mve,Thrane:2018qnx}.
If we have the data of $N$ BBH detections, $\vd = (d_1, \dots, d_N)$, then the likelihood for an inhomogeneous Poisson process is \cite{Wysocki:2018mpo,Fishbach:2018edt,Mandel:2018mve,Thrane:2018qnx}
\e\label{likelihood}
p(\vd|\vth, R) \propto R^{N} e^{-R\, \beta(\vth)} \prod_i^N 
\int \rd\vla\ p(d_i|\vla)\ p(\vla|\vth),
\q 
where $\vla \equiv \{m_1, m_2\}$.
The likelihood of an individual event $p(d_i|\vla)$ is proportional to the posterior of that event $p(\vla|d_i)$, as  the standard priors on masses for each event in \lvc\ analysis are taken to be uniform. 
We will use the publicly available posterior samples of ten BBHs \cite{TheLIGOScientific:2016pea,LIGOScientific:2018mvr} from \lvc\ observations to evaluate the integral in \Eq{likelihood}.
Meanwhile, $\beta(\vth)$ is defined as
\e 
\beta(\vth) \equiv \int \rd\vla\ VT(\vla)\ p(\vla|\vth),
\q 
in which $VT(\vla)$ is given by \Eq{VT}.
The posterior probability distribution $p(\vth, R|\vd)$ can be directly estimated by
\e\label{post} 
p(\vth, R|\vd) \propto p(\vd |\vth, R)\ p(\vth, R),
\q 
where as usual the prior distribution $p(\vth, R)$ \citep{Abbott:2016nhf,Abbott:2017vtc} is chosen to be uniform for $\vth$ parameters and log-uniform for local merger rate $R$,
\e 
p(\vth, R) \propto \frac{1}{R}.
\q 
Integrating over $R$ in \Eq{post}, it is then easy to obtain the marginalized posterior  
\e\label{post_vth} 
p(\vth|\vd) \propto \[\beta(\vth)\]^{-N} 
\prod_i^N \int \rd\vla\ p(d_i|\vla)\ p(\vla|\vth),
\q 
which has been widely used in previous population inferences \citep{Abbott:2016nhf,Abbott:2017vtc,TheLIGOScientific:2016pea,Abbott:2016drs,Fishbach:2017zga,Chen:2018rzo}.

\begin{figure}[htbp!]
	\centering
	\includegraphics[width = 0.48\textwidth]{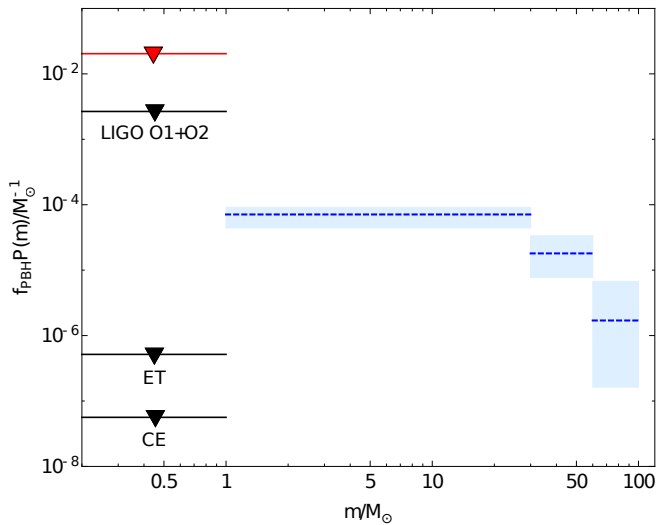}
	\caption{\label{fpbhPmdm}
		Constraints on the abundance of PBHs, $\fpbh$, in CDM.
		The blue regions with $1 \Msun \le m \le 100 \Msun$ 
		are inferred from LIGO's O1 and O2 events, 
		where the centered dashed lines are the median values and
		the shaded bars represent the $90\%$ Poisson errors.
		Four lines shown in the mass range $[0.2, 1]\Msun$ represent
		the constraints from null targeted searches of LIGO O1, 
		LIGO O1 \& O2, ET and CE, respectively.
	}
\end{figure}
Using ten BBH events from LIGO's O1 and O2 runs, we find the median value and $90\%$ equal-tailed credible intervals for the parameters $\{\vth, R\}$ to be
$P_1 = 2.1^{+0.7}_{-0.8} \times 10^{-2}$,
$P_2 = 5.4^{+4.7}_{-3.1} \times 10^{-3}$,
$P_3 = 5.1^{+15.2}_{-4.6} \times 10^{-4}$,
and $R = 308^{+193}_{-135}\, \gpcyr$, from which we also infer the fraction of PBHs in CDM to be $\fpbh = 3.3\,^{+2.3}_{-1.8} \times 10^{-3}$. 
Such an abundance of PBHs is consistent with previous estimations that $10^{-3} \lesssim \fpbh \lesssim 10^{-2}$, confirming that the dominant fraction of CDM should not originate from PBHs in the mass range of $[0.2, 100]\Msun$ \citep{Sasaki:2016jop,Ali-Haimoud:2017rtz,Raidal:2017mfl,Kocsis:2017yty,Chen:2018czv,Chen:2018rzo}. 
From now on we will investigate the possibility of detecting sub-solar mass BBHs.
Here we denote the abundance of PBHs in the mass range of $[0.2, 1]\Msun$ as 
\e 
\fpbhn \equiv \fpbh P_0 \,\Delta m_0,
\q
where $\Delta m_0 = (1 - 0.2) \Msun = 0.8 \Msun$.
As a consequence of the above analysis, it is then straightforward to infer the upper bound of $\fpbhn$ to be $\fpbhn \le 1.8 \times 10^{-3}$ by LIGO's O1 and O2 runs.
In the future, if third-generation ground-based GW detectors are in operation, the detection ability will be greatly enhanced and we have more chance to detect the sub-solar mass BBHs if they do exist.
In addition to search for the BBHs with two sub-solar mass components as \lvc\ have done, we also propose to search for the BBHs with one sub-solar mass component (with mass lying in $[0.2, 1]\Msun$) and another super-solar mass one (with mass lying in $[1, 100]\Msun$).
Using the loudest event statistic formalism [see \Eq{R90}] and the values of $\vth$ inferred from LIGO's O1 and O2 runs, $\fpbhn$ can be constrained to an unprecedented level.
Assuming no such BBHs will be detected, ET implies $\fpbhn \le 4.1 \times 10^{-7}$ while CE implies $\fpbhn \le 4.5 \times 10^{-8}$.
The results of the constraints on $\fpbh$ (and $\fpbhn$) when PBHs have a broad mass distribution are shown in \Fig{fpbhPmdm}.
Note that the red line in \Fig{fpbhPmdm} shows the upper limit of $\fpbhn \le 1.6 \times 10^{-2}$ from the null targeted search of BBHs with two sub-solar mass components, assuming that PBHs take a flat distribution in the mass range $[0.2, 1]\Msun$.

It is worthy to note that the targeted search of BBHs with a sub-solar mass and a super-solar mass components will improve the detectable limit of $\fpbh$ by an order of $\Od(10^2 \sim 10^3)$, comparing to the targeted search for BBHs with two sub-solar mass BHs as shown in \Fig{fpbh}.

\section{\label{stellar}Distinguish PBHs from ABHs by Super-Solar Mass BHs}

\begin{figure}[tbp!]
	\centering
	\includegraphics[width = 0.48\textwidth]{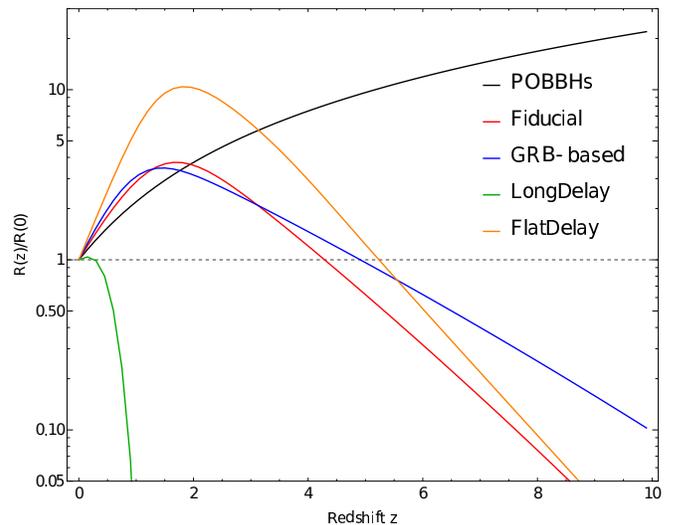}
	\caption{\label{R_z}
		Redshift distribution of the normalized merger rate, $R(z)/R(0)$,
		for the POBBHs and AOBBHs, respectively.
		For both the POBBHs and AOBBHs, we only count the BBHs with masses 
		in the range of $5\Msun \leq m_2 \leq m_1 \leq 95 \Msun$.
        We assume PBHs have a broad mass distribution of \Eq{para}, and the best-fits values are used to calculate the merger rate of POBBHs.
        See text for the details on the assumptions on different AOBBH models.
	}
\end{figure}

Besides the method of using sub-solar mass BBHs to distinguish PBHs from ABHs, there is another way by exploring the redshift evolution of the event rate of super-solar mass BBHs. 
In \cite{Chen:2018czv}, we found that the merger rate of POBBHs increases as a function of redshift $z$, namely $\mR(z) \propto t(z)^{-34/37}$, which is independent of both the abundance and mass function of PBHs. 
However, the merger rate predicted by the AOBBHs will firstly increase with $z$, then peaks at some low redshift, and lastly rapidly decreases with $z$.
\Fig{R_z} shows the merger rate of POBBHs and AOBBHs as a function of redshift $z$.
The difference between the merger rate of these two models increases at the higher redshift. 
Currently, LIGO can only observe BBHs at low redshifts (with $z < 1$), but future GW detectors such as CE and ET will be able to probe much higher redshifts (with $z \ge 10$).
In this section, we will demonstrate how well the third-generation ground-based detectors like CE and ET can be used to distinguish PBHs from ABHs according to the quite different event rate distributions at high redshifts. 

To comply with the previous studies \citep{Abbott:2017vtc,Abbott:2017xzg}, 
we restrict the component masses of BBHs to the range 
$\mmin \leq m_2 \leq m_1$
and $m_1 + m_2 \leq \mmax$, with $ \mmin = 5\Msun$ and $\mmax = 100\Msun$.
To calculate the observable events rate, we first need to know the merger rate distribution.
For PBHs, we assume they have a broad mass distribution of \Eq{para} and adopt the best-fits from \Sec{general}.
For ABHs, the merger rate is a convolution of the birthrate of ABHs $R_{\mathrm{birth}}(z,m)$ with the distribution of the time delays $P_d (t_d)$ between the formation and merger of AOBBHs (see \eg \cite{Dvorkin:2016wac})
\e\label{sBHR}
\mR_{12}(z) = \int^{t_{\mathrm{max}}}_{t_{\text{min}}}  
R_{\mathrm{birth}}(t(z)-t_d, m_1) \times P_d (t_d)\ d t_d,
\q
where $t_d$ is the time delay, and $t(z)$ is the age of the Universe at merger.
The birthrate $R_{\mathrm{birth}}$ can be estimated by \citep{Dvorkin:2016wac}
\e\label{Rbirth}
    R_{\mathrm{birth}}(t,m_{\mathrm{bh}})= \int \psi [t-\tau(m)]\, \phi(m)\, \delta(m- g_{\mathrm{bh}}^{-1}(m_{\mathrm{bh}}))\, dm,
\q
where $m_{\mathrm{bh}}$ is the mass of the remnant BH, $\tau(m)$ is the lifetime of a progenitor star of mass $m$ and $\phi(m) \propto m^{-2.35}$ is the initial mass function (IMF) \cite{Salpeter:1955it}.
We consider the \textit{WWp} model \citep{Woosley:1995ip} of BH formation in which $m$ and $m_{\mathrm{bh}}$ are related by
\e\label{mbhm}
\frac{m_{\mathrm{bh}}}{m}=A \left(\frac{m}{40\Msun} \right)^{\beta} \frac{1}{\left( \frac{Z(z)}{0.01 Z_\odot} \right)^{\gamma} +1},
\q 
where $Z(z)$ is the metallicity and an explicit functional 
form can be found in \cite{Belczynski:2016obo}. 
The parameter values in \Eq{mbhm} are $A=0.3$, $\beta =0.8$ 
and $\gamma =0.2$ \citep{Dvorkin:2016wac}.
Solving \Eq{mbhm} yields the functional form $m = g_{\mathrm{bh}}^{-1}(m_{\mathrm{bh}})$.
The star formation rate (SFR) $\psi(t)$ in \Eq{Rbirth} is given by \citep{Nagamine:2003bd}
\e
\psi(z)= k \frac{a \exp[b(z-z_{m})]}{a-b+b \exp[a(z-z_{m})]},
\q
where z is the redshift and the parameter values of $\{k, a, b, z_m\}$ are dependent on the specific AOBBH formation scenarios.
In this paper, we will consider $4$ different AOBBH models.

The first one is the \texttt{Fiducial} model in which the SFR is a fit to the observations of the luminous galaxies and the fit parameters are $k=0.178 \, \Msun \, \mathrm{yr}^{-1}\, \mathrm{Mpc}^{-3}$, $z_{m}=2.00$, $a=2.37$, $b=1.80$ as given by \cite{Vangioni:2014axa}, and the time delay distribution has the form of $P_{d} \propto t_{d}^{-1}$ with $t_{\mathrm{min}} < t_{d} < t_{\mathrm{max}}$, where the minimum delay time of a binary system to evolve until 
coalescence is set to be $t_{\mathrm{min}} = 50$\,Myr, and the maximum delay time $t_{\mathrm{max}}$ is set to the Hubble time \citep{TheLIGOScientific:2016wyq}.
This model corresponds to the \textit{classical isolated binary evolution} of AOBBH model \cite{Dvorkin:2016wac}.

The second one is the \texttt{GRB-based} model which differs from the \texttt{fiducial} model by using the SFR calibrated from the gamma ray bursts (GRB) at high redshift, and the fit parameters are given by $k=0.146 \, \Msun \, \mathrm{yr}^{-1}\, \mathrm{Mpc}^{-3}$, $z_m=1.72$, $a=2.80$, $b=2.46$ as given by \cite{Vangioni:2014axa}.

The third one is the \texttt{LongDelay} model which is identical to the \texttt{Fiducial} model but assumes a significantly longer minimum time delay $t_{\mathrm{min}} = 5$\,Gyr. The model is potentially consistent with the \textit{chemically homogeneous evolution} of rapidly rotating massive stars in very tight binaries \cite{Mandel:2015qlu}.

The last one is the \texttt{FlatDelay} model assuming a flat time delay distribution with $t_{\mathrm{min}}=50$\,Myr and $t_{\mathrm{max}}=1$\,Gyr. 
This model is inspired by the supposition that \textit{dynamical formation} of the most massive binaries is likely to happen fairly early in the history of the host environment \cite{TheLIGOScientific:2016wyq}.

\begin{figure}[tbp!]
	\centering
	\includegraphics[width = 0.48\textwidth]{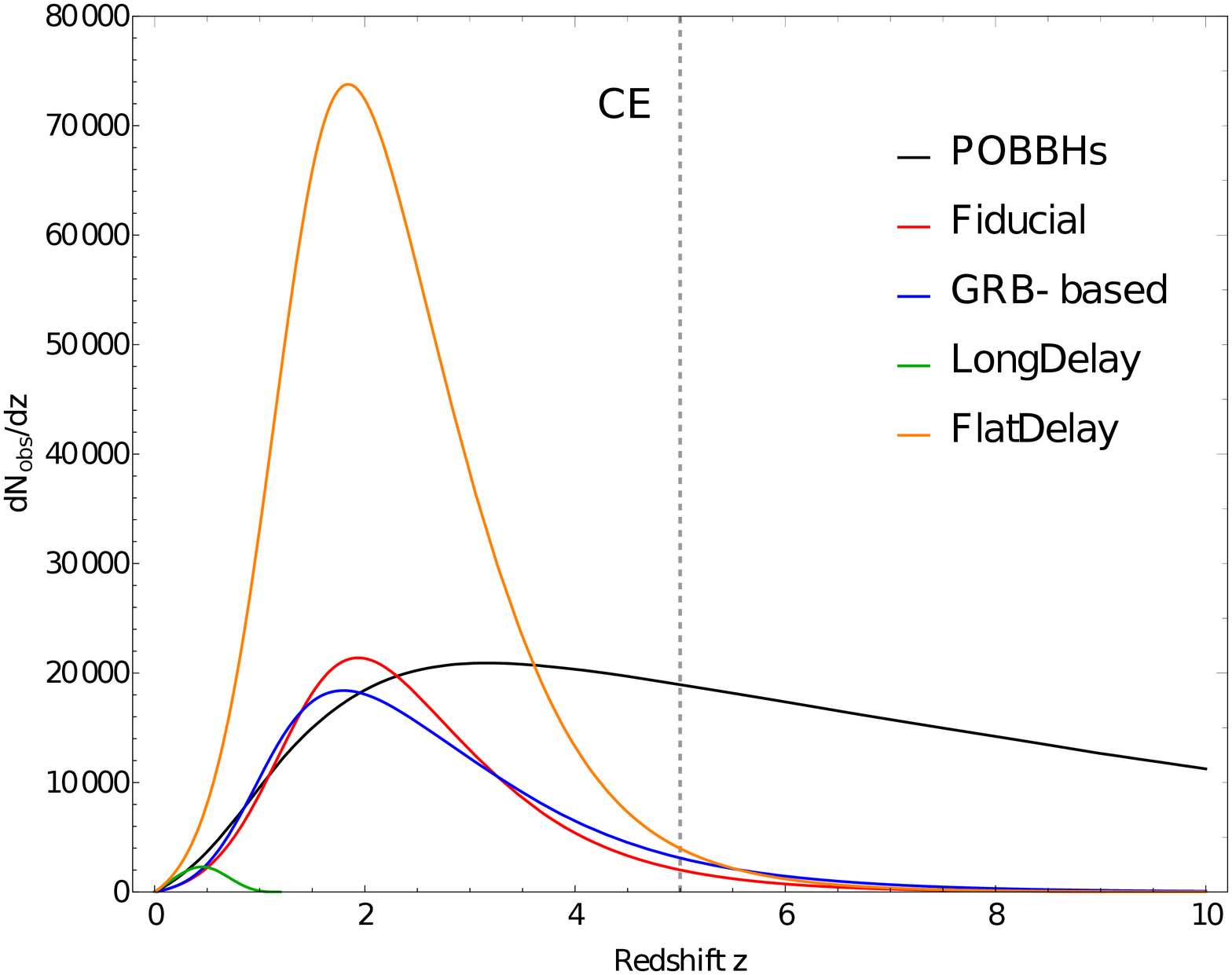}\\
	\includegraphics[width = 0.48\textwidth]{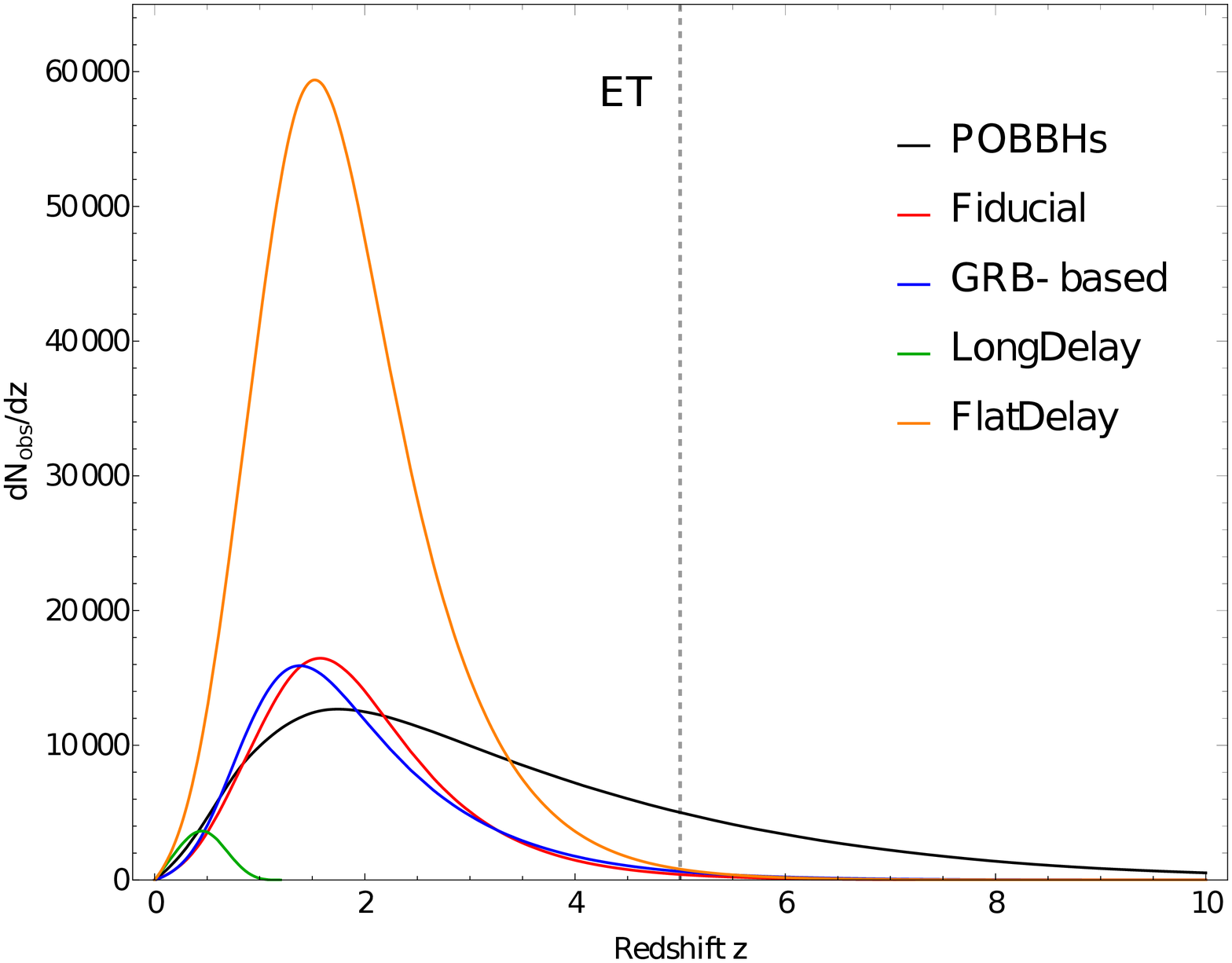}
	\caption{\label{events_z}
		Redshift distribution of the expected number of observable BBHs,
		$\rd \Nobs/\rd z$, for CE (top panel) and ET (bottom panel), respectively.
		For both the POBBHs and AOBBHs, we only count the BBHs with masses 
		in the range of $5\Msun \leq m_2 \leq m_1 \leq 95 \Msun$.
        We assume PBHs have a broad mass distribution of \Eq{para}, and the best-fits values are used to calculate the merger rate of POBBHs.
        See text for the details on the assumptions on different AOBBH models.
	}
\end{figure}

\begin{figure}[tbp!]
    \centering
    \includegraphics[width = 0.48\textwidth]{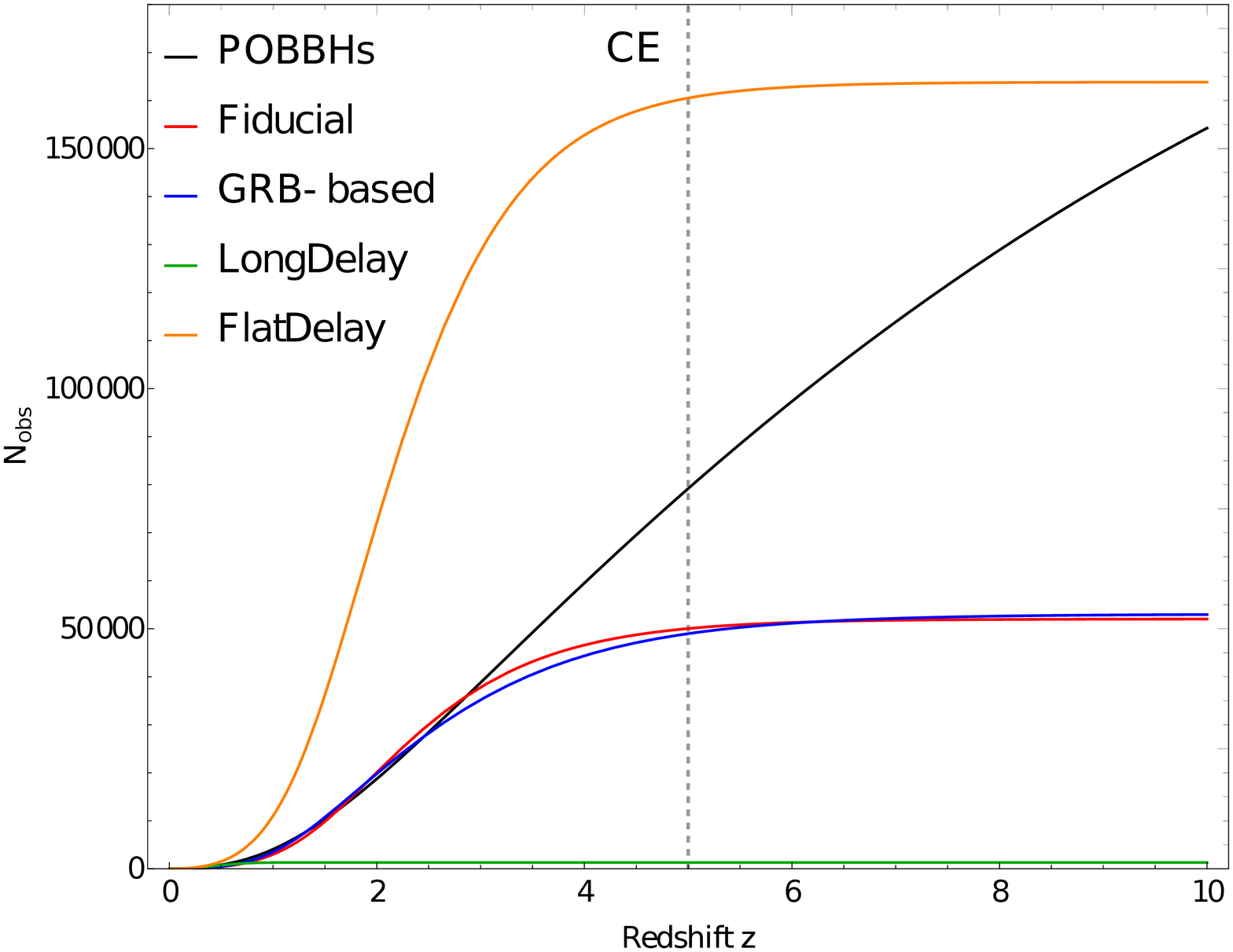}\\
    \includegraphics[width = 0.48\textwidth]{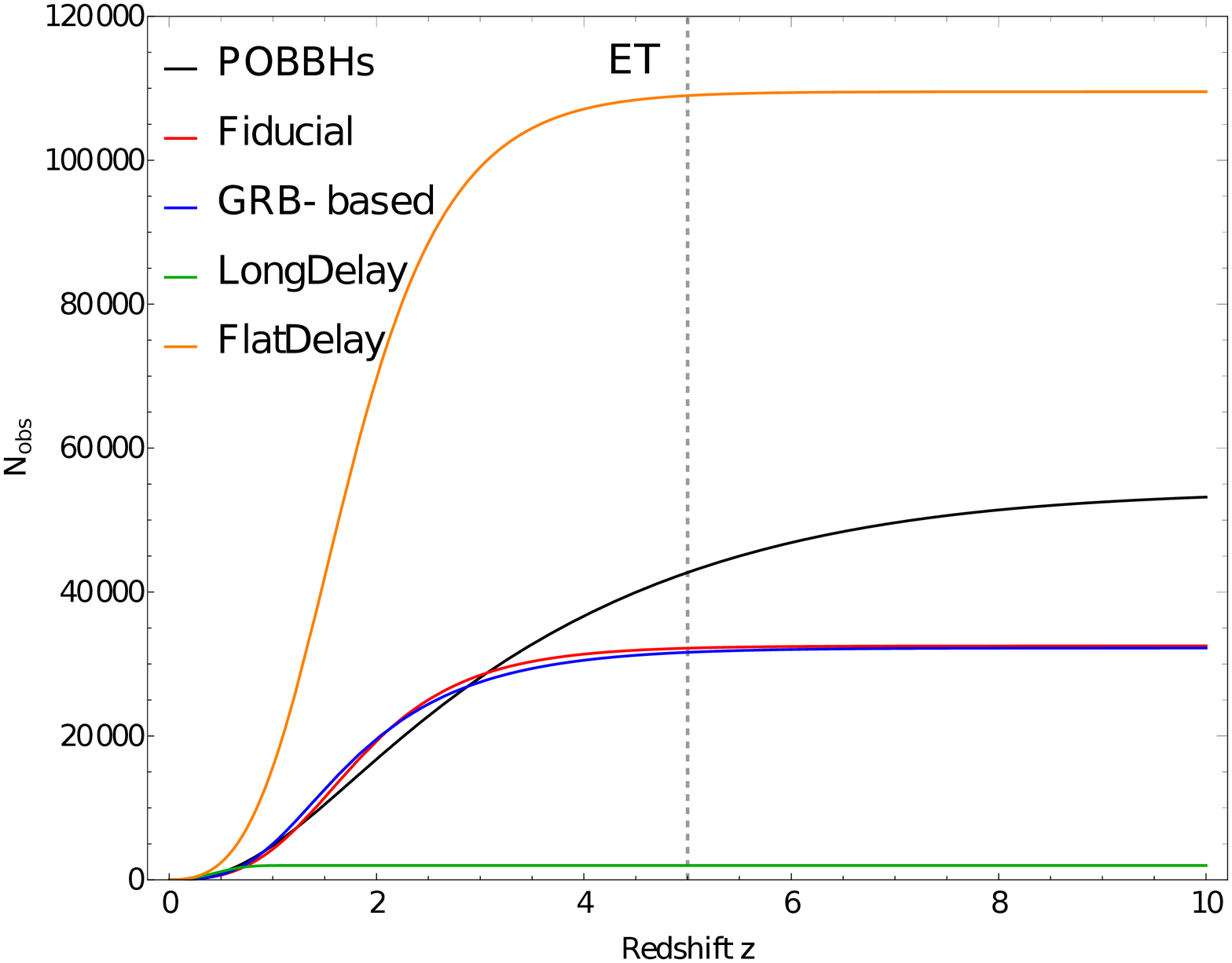}
    \caption{\label{events}
        Redshift distribution of the total number of observable BBHs,
        $\Nobs$, for CE (top panel) and ET (bottom panel), respectively.
        For both the POBBHs and AOBBHs, we only count the BBHs with masses 
        in the range of $5\Msun \leq m_2 \leq m_1 \leq 95 \Msun$.
        We assume PBHs have a broad mass distribution of \Eq{para}, and the best-fits values are used to calculate the merger rate of POBBHs.
        See text for the details on the assumptions on different AOBBH models.
    }
\end{figure}

\Fig{R_z} compares the redshift distribution of the normalized merger rate, $R(z)/R(0)$, for the POBBH and different AOBBH models, respectively.
It is shown that the merger rate of \texttt{LongDelay} model decreases quickly as redshift increases and \texttt{GRB-based} model has a relative high merger rate at high redshift comparing to the \texttt{Fiducial} and the \texttt{FlatDelay} models.
Moreover, the merger rate from all the AOBBH models decreases at high redshift while the merger rate from the POBBH model increases at high redshift.
The redshift dependent observable events number density of a GW detector can be calculated by
\e\label{event_density} 
	\frac{\rd \Nobs}{\rd z} = \int \rd m_1 \rd m_2\, \mR_{12}(z)\,
		\frac{\rd VT}{\rd z}.
\q 
Integrating over the redshift $z$ results in the total number of observable events, $\Nobs$,
\e\label{Nobs2} 
	\Nobs = \int \rd z \frac{\rd \Nobs}{\rd z}.
\q 
Note that \Eq{Nobs} is a special case of \Eq{Nobs2} when $m_1 = m_2 = m$.
\Fig{events_z} shows the expected number of observable BBHs, $\rd \Nobs/\rd z$, as a function of redshift for CE and ET, respectively.
\Fig{events} shows the total number of observable BBHs, $\Nobs$, as a function of redshift for CE and ET, respectively.
The third-generation GW detectors like CE and ET are expected to detect $\Od(10^5)$ BBH mergers each year and dig much deeper at redshit, and the fact that the redshift distribution of $\rd \Nobs/\rd z$ for POBBHs and AOBBHs are quite different from each other can be taken as a complementary tool to distinguish these two formation models of BBHs.
In particular, the contribution to the expected number of observable BBHs from high redshift (with $z>5$) for all the $4$ AOBBH models can be negligible, and hence the total number of observable events, $\Nobs$, approaches a constant number at $z>5$.
However, for the POBBHs, the contribution from the higher redshift cannot be ignored, and therefore $\Nobs$ still increases when $z>5$.
Therefore, an ``excess" of total number of observable events after redshift $z=5$ could possibly point to a population of POBBHs.

\section{\label{discuss}Summary and Discussion}

Even though several BBH coalescences have been detected by LIGO/Virgo, the origin of these BHs is still unknown. 
In this paper, we explore how well the next generation detectors, such as ET and CT, can be used to distinguish PBHs from ABHs.

Firstly, we investigate the possibility of direct detection of sub-solar mass BBHs, hence validating the existence of PBHs.
For PBHs with a monochromatic mass function, we estimate and forecast the detectable limit of $\fpbh$ from the targeted search of BBHs by LIGO, ET, and CE, respectively.
Furthermore, in order to get better sensitivity, we propose to search for the BBHs containing a sub-solar mass PBH and a super-solar mass PBH.
We predict that the abundance of PBHs in the mass range of $[0.2, 1]\Msun$ can be constrained to an order of $\Od(10^{-7})$ and $\Od(10^{-8})$ if no such BBHs are to be detected by ET and CE, respectively.

Secondly, we explore the possibility of utilizing the redshift evolution of the merger rate of super-solar mass BBHs to distinguish PBHs from ABHs.
We estimate and forecast the redshift distribution of the expected number of observable BBHs for the PBH and ABH models, respectively.
When the third generation ground-based GW detectors like CE and ET are in operation, it is expected to detect $\Od(10^5)$ BBH mergers each year and reach much deeper redshift ($z \gtrsim 10$), and the redshift distribution of detectable BBH events can serve as an alternative means to distinguish PBHs from ABHs.

Throughout this paper, we assume that all the \lvc\ BBHs originate from the same formation channel.
However, this assumption can be too oversimplified because the observed BBHs might be a mixing of POBBHs and AOBBHs.
To identify each BBH as a POBBH or AOBBH will be quite difficult in this scenario.
In addition to the mass and redshift distribution of BBHs, other information, \eg spin distribution, will also be invaluable in order to find out the progenitors of the \lvc\ BBHs.
For instance, it is expected that PBHs formed in the early universe have negligible spins \cite{Chiba:2017rvs,Mirbabayi:2019uph,DeLuca:2019buf}, while the ABHs which originated from the Population III star binaries, favor a relatively high spin distribution \cite{Kinugawa:2016ect}.

\begin{acknowledgments}
We would like to thank Lu Chen, Yun Fang, Fan Huang, Jun Li, Lang Liu, Shi Pi, 
You Wu, Yu Sang, Sai Wang, Hao Wei, Chen Yuan, and Xue Zhang 
for useful conversations. 
We acknowledge the use of HPC Cluster of ITP-CAS. 
This work is supported by grants from NSFC (grant No. 11690021, 11975019, 11947302, 11991053), the Strategic Priority Research Program of Chinese Academy of Sciences (Grant No. XDB23000000, XDA15020701), and Key Research Program of Frontier Sciences, CAS, Grant NO. ZDBS-LY-7009. 
This research has made use of data, software and/or web tools obtained 
from the Gravitational Wave Open Science Center \cite{Vallisneri:2014vxa}
(\url{https://www.gw-openscience.org}), a service of LIGO Laboratory, 
the LIGO Scientific Collaboration and the Virgo Collaboration. 
LIGO is funded by the U.S. National Science Foundation. 
Virgo is funded by the French Centre National de Recherche Scientifique (CNRS),
the Italian Istituto Nazionale della Fisica Nucleare (INFN) and the Dutch Nikhef,
with contributions by Polish and Hungarian institutes.
\end{acknowledgments}

\appendix*
\section{Constrain $\fpbh$ by SGWB}
Another way to constrain $\fpbh$ is through the SGWB \cite{Wang:2016ana} which is a superposition of the energy spectra emitted by the BBHs that are unlikely to be resolved individually.
This method differs from the one by targeted search mainly due to it also utilizes the redshift dependence of the merger rate (see \Eq{OmegaGW} below).
Notice that the targeted search is more sensitive to the local merger rate.
The energy-density spectrum of a SGWB is characterized by the dimensionless quantity \cite{Allen:1997ad}
\e\label{OmegaGW1}
	\ogw(\nu) = \frac{\nu}{\rho_{c}} \frac{\rd\rhoGW}{\rd\nu},
\q
where $\rho_{c}=3H_{0}^2/(8 \pi)$ is the critical energy density of our universe, $d\rhoGW$ is the energy density in the frequency interval $\[\nu,\, \nu + d\nu\]$, and we take the value of Hubble constant $H_0 = 67.74\, \km \sec^{-1} \Mpc^{-1}$ from Planck \citep{Ade:2015xua}.
For the binary mergers, the magnitude of a SGWB can be calculated via
\cite{Phinney:2001di,Regimbau:2008nj,Zhu:2011bd,Zhu:2012xw}
\e\label{OmegaGW}
	\ogw(\nu) = \frac{\nu}{\rho_c H_0} \int_0^{z_{\mathrm{max}}} \rd z 
		 \frac{R(z)}{(1+z)\, E(z)} \frac{\rd E_{\mathrm{GW}}}{\rd \nu_s},
\q
where $\nu_s = (1+z)\, \nu$ is the frequency in source-frame, $E(z)=\sqrt{\Om_r \(1+z\)^4 + \Om_m (1+z)^3+\Omega_{\Lambda}}$ accounts for the evolution of our universe, and the factor $(1+z)$ in the denominator of \Eq{OmegaGW} converts the merger rate, $R(z)$, from source frame to detector frame. 
We adopt the best-fit results from Planck \citep{Ade:2015xua} that $\Om_r = 9.15 \times 10^{-5}$, $\Om_m = 0.3089$, and $\Om_\Lambda = 1 - \Om_m - \Om_r$.
The cutoff redshift is chosen to be $z_{\mathrm{max}} = \nu_3/\nu - 1$ \cite{Wang:2016ana}, in which $\nu_3$ is given by \Eq{dEdnu} below.
Furthermore, the energy spectrum $\rd E_{\mathrm{GW}}/\rd \nu_{s}$, emitted by an individual BBH with equal component masses $m_1 = m_2 = m$, is approximated by \cite{Cutler:1993vq,Chernoff:1993th,Zhu:2011bd}
\e\label{dEdnu} 
\frac{\rd E_{\mathrm{GW}}}{\rd \nu_s} = \frac{\pi^{2/3} M^{5/3} \eta}{3} 
\begin{cases}
	\nu_s^{-1/3}, &\hspace{-2mm} \nu_s<\nu_1,\\
	\frac{\nu_s}{\nu_1} \nu^{-1/3}, &\hspace{-2mm} \nu_1 \leq \nu_s < \nu_2,\\
	\frac{\nu_s^2}{\nu_1 \nu_2^{4/3}} \frac{\nu_4^4}{\(4\(\nu_s-\nu_2\)^2 + \nu_4^2\)^2}, 
	&\hspace{-2mm} \nu_2 \leq \nu_s < \nu_3,
\end{cases}
\q
where $M = m_1 + m_2 = 2\, m$ is the total mass of the BBH, $\eta = m_1 m_2 / M^2 = 1/4$, and $\nu_i = \(a_i \eta^2 + b_i \eta + c_i\)/\(\pi M\)$ with $i = \{1, 2, 3\}$. 
The coefficients $a_i$, $b_i$ and $c_i$ are presented in Table~I of \cite{Ajith:2007kx}. 
Again, all masses are measured in the source frame and units of $\Msun$.

For a network of $n$ individual detectors, the SNR $\rho$ for measuring the SGWB with an observation time $\Tobs$ is given by \citep{Allen:1997ad,Thrane:2013oya}
\e\label{SNR}
	\rho = \sqrt{2 \Tobs} \[ \int \rd\nu \sum_{I=1}^n \sum_{J>I}^n
		\frac{\Gamma^2_{IJ}(\nu)\, S_h^2(\nu)}
		{P_{nI}(\nu)\, P_{nJ}(\nu)} \]^{1/2},
\q
where $P_{nI}(\nu)$ is the auto  power spectral density for the noise in detector $I$ and $\Gamma_{IJ}(\nu)$ is the overlap reduction function \cite{Christensen:1992wi,Flanagan:1993ix}.
In \Eq{SNR}, $S_h$ is the strain power density spectrum of a SGWB, which is related to $\ogw$ through \cite{Thrane:2013oya}
\e 
	S_h(\nu) = \frac{3 H_0^2}{2 \pi^2} \frac{\ogw(\nu)}{\nu^3}.
\q 
Here we set $\rho = 1$, which corresponds to $1\s$ confidential level, as the criterion for the detection of SGWBs.

\bibliography{./bibfile}

\end{document}